\documentclass[12pt]{article}
\usepackage{amsmath}
\usepackage{amssymb} 
\usepackage{epsfig}
\usepackage{amsthm}
\usepackage{stmaryrd}
\usepackage{bm}
\usepackage{hyperref}
\usepackage{eufrak}
\usepackage{upgreek}

\numberwithin{equation}{section}
\numberwithin{figure}{section}

\def\eq#1{(\ref{eq:#1})}
\def\lineup{\!\!\!\!\!\!\!\!&&}

\def\d{\partial}
\def\eps{\epsilon}

\def\deg{\mathrm{deg}}
\def\Mb{\overline{M}}

\def\M{{\bf M}}
\def\Q{{\bf Q}}
\def\n{{\bm \upeta}}
\def\nb{\overline{\bm \upeta}}
\def\b{{\bf b}}
\def\c{{\bf c}}
\def\L{{\bf L}}
\def\mmu{{\bm \upmu}}
\def\l{\lambda}
\def\lb{\overline{\lambda}}
\def\ll{{\bm \uplambda}}
\def\llb{\overline{\bm \uplambda}}
\def\w{\wedge}

\begin{document}

\begin{titlepage}
\hfill LMU-ASC 09/14

\begin{center}

\vskip 1.5cm {\large \bf{NS-NS Sector of Closed Superstring Field Theory}}

\vskip 1.5cm

{\large Theodore Erler\footnote{tchovi@gmail.com}, Sebastian 
Konopka\footnote{sebastian.konopka@physik.uni-muenchen.de}, 
Ivo Sachs\footnote{ivo.sachs@physik.uni-muenchen.de}}

\vskip 1.0cm

{\it Arnold Sommerfeld Center, Ludwig-Maximilians University,}\\
{\it Theresienstrasse 37, D-80333, Munich, Germany}
\vskip 1.5cm

{\bf Abstract}

\end{center}

We give a construction for a general class of vertices in superstring field 
theory which include integration over bosonic moduli as well as the required 
picture changing insertions. We apply this procedure to find a covariant 
action for the NS-NS sector of Type II closed superstring field theory.

\noindent

\noindent
\medskip

\end{titlepage}

\tableofcontents

\section{Introduction}

Though bosonic string field theory has been well-understood since the 
mid 90's \cite{Witten,ZwiebachClosed,ZwOpCl,Zwiebach}, superstring 
field theory remains largely mysterious. In some cases it is possible 
to find elegant formulations utilizing the large 
Hilbert space \cite{Berkovits,BerkRamond,heterotic1,heterotic,hetRamond}, 
but it seems difficult to push beyond tree level 
\cite{BerkPert,BerkPertBerk,Torii1,Torii2} and the 
presumed geometrical underpinning of the theory in terms of the supermoduli 
space remains obscure. A somewhat old-fashioned alternative 
\cite{WittenSuper} is to formulate superstring field theory using fields in 
the small Hilbert space. A well known complication, however, is 
that one needs a prescription for inserting picture changing operators 
into the action. This requires an apparently endless sequence 
of choices, and while limited work in this direction exists 
\cite{Sen,Yeh,Muenster}, it has not produced a compelling and fully explicit 
action. 

Recent progress on this problem for the open superstring was reported in 
\cite{WittenSS}, inspired by studies of gauge fixing in Berkovits' open 
superstring field theory \cite{OkWit}. The basic insight of \cite{WittenSS}
is that the multi-string products of open superstring field theory can be 
constructed by passing to the large Hilbert space and constructing 
a particular finite gauge transformation through the space of 
$A_\infty$ structures. The result is an explicit action for open superstring 
field theory which automatically satisfies the classical BV master 
equation. In this paper we generalize these results to define classical 
actions for the NS sectors of all open and closed superstring field theories. 
Of particular interest is the NS-NS sector of Type II closed superstring field 
theory, for which a construction in the large Hilbert space 
appears difficult \cite{BartonTypeII}.\footnote{A recent
proposal for Type II closed superstring field theory in the large Hilbert 
space appears in \cite{Matsunaga}. Interestingly, however, picture changing 
operators still appear to be needed in the action.} The main technical 
obstacle for us 
will be learning how to accommodate vertices which include integration
over bosonic moduli, and for the NS-NS superstring, how to insert 
additional picture changing operators for the rightmoving sector. These 
results lay the groundwork for serious consideration of the Ramond sector and 
quantization of superstring field theory. This is of particular interest 
in the context of recent efforts to obtain a more complete understanding 
of superstring perturbation theory 
\cite{Belopolsky,WittenPert,WittenDonagi,SenRen1,SenRen2}.

This paper is organized as follows. In section \ref{sec:AinfLinf} we review
the algebraic formulation of open and closed string field theory in terms 
of $A_\infty$ and $L_\infty$ algebras, with an emphasis on the coalgebra 
description. This mathematical language gives a compact and convenient 
notation for expressing various multi-string products and their 
interrelation. In section \ref{sec:Wstubs} we revisit Witten's open 
superstring field theory in the $-1$ picture \cite{WittenSuper}, but 
generalizing \cite{WittenSS}, we allow vertices which include integration 
over bosonic moduli as well as the required picture changing insertions. 
We find that the multi-string products can be derived from a recursion 
involving a two-dimensional array of products of intermediate picture number. 
The recursion emerges from the solution to a pair of differential equations
which follow uniquely from two assumptions: that the products 
are derived by gauge transformation through the space of $A_\infty$ 
structures, and that the gauge transformation is defined 
in the large Hilbert space. In section \ref{sec:het}, 
we explain how this construction generalizes (with little effort) to the 
NS sector of heterotic string field theory. In section \ref{sec:NSNS} we 
consider the NS-NS sector of Type II closed superstring field theory. 
We give one construction which 
defines the products by applying the open string recursion of section 
\ref{sec:Wstubs} twice, first to get the correct picture in the leftmoving 
sector and and again to get the correct picture in the rightmoving sector. 
This construction however treats the left and rightmoving sectors 
asymmetrically. We therefore provide a second, more nontrivial construction 
which preserves symmetry between left and righmovers at every stage in the 
recursion. We end with some conclusions.

\section{$A_\infty$ and $L_\infty$ Algebras}
\label{sec:AinfLinf}

Here we review the algebraic formulation of open and closed string field
theory in the language on $A_\infty$ and $L_\infty$ algebras. For the 
$A_\infty$ case the discussion basically repeats section 4 of 
\cite{WittenSS}. For more mathematical discussion see \cite{Kajiura} 
for $A_\infty$ and \cite{Linf,Linf2} for $L_\infty$.

\subsection{$A_\infty$}

Let's start with the $A_\infty$ case. Here the basic objects are 
multi-products $b_n$ on a $\mathbb{Z}_2$-graded vector space $\mathcal{H}$
\begin{equation}b_n(\Psi_1,...,\Psi_n)\in\mathcal{H},
\ \ \ \ \Psi_i\in\mathcal{H},\label{eq:prod_b}
\end{equation}
which have no particular symmetry upon interchange of the arguments. For us,
$\mathcal{H}$ is the open string state space and the $\mathbb{Z}_2$ grading,
called {\it degree}, is Grassmann parity plus one. The product 
$b_n$ defines a linear map from the $n$-fold tensor product of 
$\mathcal{H}$ into $\mathcal{H}$:
\begin{equation}b_n:\mathcal{H}^{\otimes n}\to\mathcal{H}.\end{equation}
If we have a state in $\mathcal{H}^{\otimes n}$ of the form
\begin{equation}\Psi_1\otimes\Psi_2\otimes...\otimes\Psi_n\in
\mathcal{H}^{\otimes n},\label{eq:Hn_basis}\end{equation}
then $b_n$ acts on such a state as 
\begin{equation}b_n(\Psi_1\otimes...\otimes\Psi_n)=
b_n(\Psi_1,...,\Psi_n),\end{equation}
where the right hand side is the multi-string product as written in 
\eq{prod_b}. Since the states \eq{Hn_basis} form a basis, this equation 
defines the action of $b_n$ on the whole tensor product space.

Now suppose we have two linear maps
\begin{eqnarray}\lineup 
A:\mathcal{H}^{\otimes k}\to\mathcal{H}^{\otimes l},\nonumber\\
\lineup 
B:\mathcal{H}^{\otimes m}\to\mathcal{H}^{\otimes n}.
\end{eqnarray}
We will find it useful to define the tensor product map:
\begin{equation}A\otimes B:\mathcal{H}^{\otimes k+m}
\to\mathcal{H}^{\otimes l+n}.\end{equation}
Applying this to a state of the form \eq{Hn_basis} gives 
\begin{eqnarray}
A\otimes B(\Psi_1\otimes...\otimes\Psi_k\otimes\Psi_{k+1}
\otimes...\otimes\Psi_{k+m}) \lineup 
= (-1)^{\deg{B}(\deg{\Psi_1}+...+\deg{\Psi_k})}
\nonumber\\
\lineup\ \ \ \ \times
A(\Psi_1\otimes...\otimes\Psi_k)\otimes B(\Psi_{k+1}
\otimes...\otimes\Psi_{k+m}).\nonumber\\\label{eq:tenmap}
\end{eqnarray}
There may be a sign from commuting $B$ past the first $k$ states. 
We are particularly interested in tensor products of $b_n$ with 
the identity map on $\mathcal{H}$, which we denote $\mathbb{I}$.

With these preparations, we can define a natural action of the $n$-string 
product $b_n$ on the tensor algebra of $\mathcal{H}$:\footnote{In our case, 
we should identify $\mathcal{H}^{\otimes 0}=\mathbb{C}$.}
\begin{equation}\b_n:T\mathcal{H}\to T\mathcal{H},\ \ \ \ \ 
T\mathcal{H} = 
\mathcal{H}^{\otimes 0}
\oplus\mathcal{H}\oplus\mathcal{H}^{\otimes 2}\oplus...\ .\end{equation}
The tensor algebra has a natural coalgebra structure, on which $\b_n$ acts 
as a coderivation (see, for example, \cite{Kajiura}). We will usually 
indicate the coderivation corresponding to an $n$-string product with boldface.
The coderivation $\b_n$ can be defined by its action on each 
$\mathcal{H}^{\otimes N}$ component of the tensor algebra. If it acts on 
$\mathcal{H}^{\otimes N\geq n}$, we have 
\begin{equation}\b_n\Psi \equiv \sum_{k=0}^{N-n}\mathbb{I}^{\otimes N-n-k}
\otimes b_n\otimes \mathbb{I}^k\Psi,\ \ \ \ \ 
\Psi\in\mathcal{H}^{\otimes N\geq n}\subset T\mathcal{H}.\end{equation}
If it acts on the $\mathcal{H}^{\otimes N<n}$ component, by definition $\b_n$
vanishes. One can check that the commutator\footnote{Commutators in this 
paper are always graded with respect to degree.} of two coderivations is 
also a coderivation. For example, if $\b_m$ and $\c_n$
are coderivations derived from the multi-string products
\begin{eqnarray}
\lineup b_m:\mathcal{H}^{\otimes m}\to\mathcal{H},\nonumber\\
\lineup c_n:\mathcal{H}^{\otimes n}\to \mathcal{H},
\end{eqnarray}
then the commutator $[\b_m,\c_n]$ is a coderivation derived from the $m+n-1$ 
string product,
\begin{equation}
[b_m,c_n]\equiv b_m\sum_{k=0}^{m-1}\mathbb{I}^{\otimes m-1-k}\otimes 
c_n\otimes\mathbb{I}^{\otimes k}-(-1)^{\deg(b_m)\deg(c_n)}
c_n\sum_{k=0}^{n-1}\mathbb{I}^{\otimes n-1-k}\otimes 
b_m\otimes\mathbb{I}^{\otimes k}.
\label{eq:cocom}\end{equation}
This means that multi-string products in open string field theory, packaged
in the form of coderivations, naturally define a graded Lie algebra. This 
fact is very useful for simplifying the expression of the $A_\infty$ 
relations.

Open string field theory is defined by a sequence of multi-string products of 
odd degree satisfying the relations of a cyclic $A_\infty$ algebra. We denote
these products
\begin{equation}M_1\!=\!Q,\ M_2,\ M_3,\ M_4,\ ...\ ,\end{equation}
where $Q$ is the BRST operator and 
\begin{equation}M_n:\mathcal{H}^{\otimes n}\to \mathcal{H}.\end{equation}
The $A_\infty$ relations imply that the BRST variation of the $n$th product 
$M_n$ is related to sums of compositions of lower products $M_{k<n}$. 
This is most conveniently expressed using coderivations:
\begin{equation}[\M_1,\M_n]+[\M_2,\M_{n-1}]+...+[\M_{n-1},\M_2]+[\M_n,\M_1]=0.
\label{eq:Ainf}\end{equation}
The first and last terms represent the BRST variation of $M_n$. For example,
the fact that $Q$ is a derivation of the 2-product is expressed by the 
equation,
\begin{equation}[\Q,\M_2]=0.\end{equation}
Using \eq{cocom}, this implies
\begin{equation}QM_2 + M_2(Q\otimes \mathbb{I}+\mathbb{I}\otimes Q)=0,
\end{equation}
and acting on a pair of states $\Psi_1\otimes\Psi_2$ gives
\begin{equation}QM_2(\Psi_1,\Psi_2)+M_2(Q\Psi_1,\Psi_2)+(-1)^{\deg(\Psi_1)}
M_2(\Psi_1,Q\Psi_2)=0,\end{equation}
which is the familiar expression of the fact that $Q$ is a derivation 
(recalling that $M_2$ has odd degree.)
To write the action, we need one more ingredient: a symplectic form
\begin{equation}\langle \omega|:\mathcal{H}^{\otimes 2}\to \mathbb{C}.
\end{equation}
Writing $\langle\omega|\Psi_1\otimes\Psi_2=\omega(\Psi_1,\Psi_2)$, the 
symplectic form is related to the BPZ inner product through
\begin{equation}\omega(\Psi_1,\Psi_2)=(-1)^{\deg(\Psi_1)}\langle 
\Psi_1,\Psi_2\rangle,\end{equation}
and is graded antisymmetric:
\begin{equation}\omega(\Psi_1,\Psi_2)=-(-1)^{\deg(\Psi_1)\deg(\Psi_2)}
\omega(\Psi_2,\Psi_1).\end{equation}
Gauge invariance requires that $n$-string products are BPZ odd:
\begin{equation}\langle \omega|\mathbb{I}\otimes M_n 
= -\langle\omega|M_n\otimes\mathbb{I},\end{equation}
so that they give rise to cyclic vertices (in this case the products define 
a so-called {\it cyclic} $A_\infty$ algebra). Then we can write a gauge 
invariant action
\begin{equation}S=\sum_{n=0}^\infty\frac{1}{n+2}
\omega(\Psi,M_{n+1}(\underbrace{\Psi,...,\Psi}_{n+1\ \mathrm{times}})).
\end{equation}

\subsection{$L_\infty$}

Now let's discuss the $L_\infty$ case. The basic objects are 
multi-products $b_n$ on a $\mathbb{Z}_2$-graded vector space $\mathcal{H}$
\begin{equation}b_n(\Phi_1,...,\Phi_n)\in\mathcal{H},
\ \ \ \ \Phi_i\in\mathcal{H},
\label{eq:bc}\end{equation}
which are {\it graded symmetric} upon interchange of the arguments. For us, 
$\mathcal{H}$ is the closed string state space, and the $\mathbb{Z}_2$ 
grading, called {\it degree}, is identical to Grassmann parity (unlike for 
the open string, where degree is identified with Grassmann parity plus one.). 
Since the products are (graded) symmetric upon interchange of inputs, they 
naturally act on a symmetrized tensor algebra. We will denote the 
symmetrized tensor product with a wedge $\w$. It satisfies
\begin{equation}
\Phi_1\w\Phi_2 = (-1)^{\deg(\Phi_1)\deg(\Phi_2)}\Phi_2\w\Phi_1,
\ \ \ \Phi_1\w(\Phi_2\w\Phi_3)=(\Phi_1\w\Phi_2)\w\Phi_3.
\end{equation}
The wedge product is related to the tensor product through the formula
\begin{equation}
\Phi_1\w\Phi_2\w...\w\Phi_n=\sum_\sigma(-1)^{\eps(\sigma)} 
\Phi_{\sigma(1)}\otimes\Phi_{\sigma(2)}\otimes
...\otimes\Phi_{\sigma(n)}, \ \ \ \ \ \Phi_i\in\mathcal{H}.
\label{eq:Hwn_basis}\end{equation}
The sum is over all distinct permutations $\sigma$ of $1,...,n$, and the sign
$(-1)^{\eps(\sigma)}$ is the obvious sign obtained by moving 
$\Phi_1,\Phi_2,...,\Phi_n$ past each other into the order prescribed by 
$\sigma$. Note that if some of the factors in the wedge product are the 
identical, some permutations in the sum may produce an identical term, which 
effectively produces a $k!$ for $k$ degree even identical factors (degree odd 
identical factors vanish when taking the wedge product). With these 
definitions, the closed string product $b_n$ can be seen as a linear map from 
the $n$-fold wedge product of $\mathcal{H}$ into $\mathcal{H}$:
\begin{equation}b_n:\mathcal{H}^{\w n}\to\mathcal{H}.\end{equation}
Acting on a state of the form \eq{Hwn_basis},
\begin{equation}b_n(\Phi_1\w...\w\Phi_n) = b_n(\Phi_1,...,\Phi_n),
\label{eq:actS}\end{equation}
where the right hand side is the $n$ string product as denoted in \eq{bc}. 
Since the states \eq{Hwn_basis} form a basis, this defines the action of 
$b_n$ on all states in $\mathcal{H}^{\w n}$.

We can define the wedge product between linear maps in a similar way as 
between states: We replace wedge products with tensor products and sum over 
permutations, as in \eq{Hwn_basis}. Therefore, the wedge product of linear 
maps is implicitly defined by the tensor product of linear maps, via 
\eq{tenmap}. While this seems natural, expanding multiple wedge products
out into tensor products is usually cumbersome. However, the net result 
is simple. Suppose we have two linear maps between symmetrized 
tensor products of 
$\mathcal{H}$:
\begin{eqnarray}\lineup 
A:\mathcal{H}^{\w k}\to\mathcal{H}^{\w l},\nonumber\\
\lineup 
B:\mathcal{H}^{\w m}\to\mathcal{H}^{\w n}.
\end{eqnarray}
Their wedge product defines a map 
\begin{equation}A\w B:\mathcal{H}^{\w k+m}
\to\mathcal{H}^{\w l+n}.\end{equation}
On states of the form \eq{Hwn_basis}, $A\w B$ acts as 
\begin{equation}
A\w B(\Phi_1\w\Phi_2\w...\w\Phi_{k+m})
= \sum_{\sigma}\,\!'\, (-1)^{\eps(\sigma)}
A(\Phi_{\sigma(1)}\w...\w\Phi_{\sigma(k)})\w B(\Phi_{\sigma(k+1)}
\w...\w\Phi_{\sigma(k+m)}),\label{eq:wact}
\end{equation}
where $\sigma$ is a permutation of $1,...,k+m$, and $\Sigma'$ means
that we sum only over permutations which change the inputs of $A$ and $B$. 
(Permutations which only move around inputs of $A$ and $B$ produce the same 
terms, and are only counted once). The sign $\eps(\sigma)$ is the sign 
obtained from moving the $\Phi_i$s past each other and past $B$ to obtain 
the ordering required by $\sigma$. For example, let's consider wedge products
of the identity map, where potentially confusing symmetry factors arise. Act
$\mathbb{I}\w\mathbb{I}$ on a pair of states using \eq{wact}:
\begin{eqnarray}
\mathbb{I}\w\mathbb{I}(\Phi_1\w\Phi_2)
\lineup=\mathbb{I}(\Phi_1)\w\mathbb{I}(\Phi_2)+(-1)^{\deg(\Phi_1)\deg(\Phi_2)}
\mathbb{I}(\Phi_2)\w\mathbb{I}(\Phi_1),\nonumber\\
\lineup=2\Phi_1\w\Phi_2.
\end{eqnarray}
Here we find a factor of two because there are two permutations of 
$\Phi_1,\Phi_2$ which switch entries between the first and second maps.
Alternatively, we can compute this by expanding in tensor products:
\begin{eqnarray}
\mathbb{I}\w\mathbb{I}(\Phi_1\w\Phi_2)
\lineup
=(\mathbb{I}\otimes\mathbb{I}+\mathbb{I}\otimes\mathbb{I})
\Big(\Phi_1\otimes\Phi_2
+(-1)^{\deg(\Phi_1)\deg(\Phi_2)}\Phi_2\otimes\Phi_1\Big),\nonumber\\
\lineup 
=2(\Phi_1\otimes\Phi_2+(-1)^{\deg(\Phi_1)\deg(\Phi_2)}\Phi_2\otimes\Phi_1),
\nonumber\\
\lineup
=2\Phi_1\w\Phi_2.
\end{eqnarray}
Here the factor of two comes because there are two ways to arrange the first
and second identity map (which happen to be identical). In this way, it is
easy to see that the identity operator on $\mathcal{H}^{\w n}$ is given by
\begin{equation}\mathbb{I}_n\equiv\frac{1}{n!}
\underbrace{\mathbb{I}\w...\w\mathbb{I}}_{n\ \mathrm{times}} = 
\underbrace{\mathbb{I}\otimes...\otimes\mathbb{I}}_{n\ \mathrm{times}}.
\end{equation}
The inverse factor of $n!$ is needed to cancel the $n!$ over-counting of 
identical permutations of $\mathbb{I}$.

With these preparations, we can lift the closed string product $b_n$ to a
coderivation on the symmetrized tensor algebra:\footnote{We identify 
$\mathcal{H}^{\w 0}=\mathbb{C}$.}
\begin{equation}\b_n:S\mathcal{H}\to S\mathcal{H},\ \ \ \ \ 
S\mathcal{H}=
\mathcal{H}^{\w 0}\oplus\mathcal{H}\oplus\mathcal{H}^{\w 2}\oplus...\ .
\end{equation}
On the $\mathcal{H}^{\w N\geq n}$ component of the symmetrized tensor algebra, 
$\b_n$ acts as 
\begin{equation}\b_n\Phi\equiv (b_n\w\mathbb{I}_{N-n})\Phi,\ \ \ \ \ \ \Phi\in
\mathcal{H}^{\w N\geq n} \subset S\mathcal{H},\end{equation}
and on the $\mathcal{H}^{\w N<n}$ component $\b_n$ vanishes. If $\b_m$ and 
$\c_n$ are coderivations derived from the products
\begin{eqnarray}
\lineup b_m:\mathcal{H}^{\w m}\to\mathcal{H},\nonumber\\
\lineup c_n:\mathcal{H}^{\w n}\to \mathcal{H},
\end{eqnarray}
then the commutator $[\b_m,\c_n]$ is a coderivation derived from the 
$m+n-1$-string product,
\begin{equation}
[b_m,c_n]\equiv b_m(c_n\w \mathbb{I}_{m-1})-(-1)^{\deg(b_m)\deg(c_n)}
c_n(b_m\w\mathbb{I}_{n-1}).
\label{eq:cowcom}\end{equation}
This means that, when described as coderivations on the symmetrized tensor 
algebra, the products of closed string field theory naturally define a graded 
Lie algebra.

Closed string field theory is defined by a sequence of multi-string products 
of odd degree satisfying the relations of a cyclic $L_\infty$ algebra. We 
denote these products
\begin{equation}L_1\!=\!Q,\ L_2,\ L_3,\ L_4,\ ...\ ,\end{equation}
where
\begin{equation}L_n:\mathcal{H}^{\w n}\to \mathcal{H}.\end{equation}
The $L_\infty$ relations imply that the BRST variation of the $n$th closed 
string product $L_n$ is related to sums of compositions of lower products 
$L_{k<n}$. In fact, expressed using coderivations, the $L_\infty$ relations 
have the same formal structure as the $A_\infty$ relations:
\begin{equation}[\L_1,\L_n]+[\L_2,\L_{n-1}]+...+[\L_{n-1},\L_2]+[\L_n,\L_1]=0.
\end{equation}
What makes these relations different is the $\L_n$s act on the 
symmetrized tensor algebra, rather than the tensor algebra as for the open 
string. Consider for example the third $L_\infty$ relation,
\begin{equation}[\Q,\L_3]+\frac{1}{2}[\L_2,\L_2]=0,\end{equation}
which should characterize the failure of the Jacobi identity for 
$L_2$ in terms of the BRST variation of $L_3$. To write this identity 
directly in terms of the products, use \eq{cowcom}:
\begin{equation}QL_3 + L_3(Q\w \mathbb{I}_2)+L_2(L_2\w\mathbb{I})=0.
\label{eq:Jacobi}\end{equation}
Acting on a wedge product of three states, according to \eq{wact} we must 
sum over distinct permutations of the states on the inputs. With  
\eq{actS}, this gives a somewhat lengthy expression:
\begin{eqnarray}
0\lineup =QL_3(\Phi_1,\Phi_2,\Phi_3)+L_3(Q\Phi_1,\Phi_2,\Phi_3)
+(-1)^{\deg(\Phi_1)(\deg(\Phi_2)+\deg(\Phi_3))}
L_3(Q\Phi_2,\Phi_3,\Phi_1)\nonumber\\
\lineup\ \ \ 
+(-1)^{\deg(\Phi_3)(\deg(\Phi_1)+\deg(\Phi_2))}L_3(Q\Phi_3,\Phi_1,\Phi_2)
\nonumber\\ 
\lineup\ \ \ 
+L_2(L_2(\Phi_1,\Phi_2),\Phi_3)+(-1)^{\deg(\Phi_3)(\deg(\Phi_1)+\deg(\Phi_2))}
L_2(L_2(\Phi_3,\Phi_1),\Phi_2)\nonumber\\
\lineup\ \ \ 
+(-1)^{\deg(\Phi_1)(\deg(\Phi_2)+\deg(\Phi_3))}
L_2(L_2(\Phi_2,\Phi_3),\Phi_1).
\end{eqnarray}
The first four terms represent the BRST variation of $L_3$, and the last three 
terms represent the Jacobiator computed from $L_2$.

To write the action, we need a symplectic form for closed strings:
\begin{equation}\langle \omega|:\mathcal{H}^{\otimes 2}\to \mathbb{C}.
\end{equation}
Note that $\langle \omega|$ acts on a tensor product of two 
closed string states (rather than the wedge product, which would vanish by 
symmetry). Writing 
$\langle\omega|\Phi_1\otimes \Phi_2=\omega(\Phi_1,\Phi_2)$, the symplectic 
form is related to the closed string inner product through
\begin{equation}\omega(\Phi_1,\Phi_2)=(-1)^{\deg(\Phi_1)}\langle 
\Phi_1,c_0^- \Phi_2\rangle,\label{eq:symp_cl}\end{equation}
where $c_0^-\equiv c_0-\overline{c}_0$.\footnote{The BPZ inner product 
\begin{equation}\langle \Phi_1.\Phi_2\rangle = \langle 
I\circ\mathcal{V}_{\Phi_1}(0)\mathcal{V}_{\Phi_2}(0)\rangle\end{equation}
is conventionally defined with the conformal map $I(z)=1/z$ for closed 
strings.} Closed string fields are assumed to satisfy the constraints
\begin{eqnarray}
b_0^-\Phi \lineup = 0,\ \ \ \ b_0^-\equiv b_0-\overline{b}_0,\nonumber\\
L_0^-\Phi \lineup = 0,\ \ \ \ L_0^-\equiv L_0 - \overline{L}_0.
\label{eq:level}\end{eqnarray}
With these conventions the symplectic form is graded 
antisymmetric:\footnote{The extra sign in front of the closed string 
inner product in \eq{symp_cl} was chosen to ensure graded antisymmetry of 
the symplectic form. Without the sign, the closed string inner product itself 
has the symmetry of an odd symplectic form, like the antibracket. This symmetry
however is somewhat awkward to describe in the tensor algebra language. 
Note that, with our choice of symplectic form, permutation symmetry of the 
vertices produces signs from moving fields through the products $L_n$.}
\begin{equation}\omega(\Phi_1,\Phi_2)=-(-1)^{\deg(\Phi_1)\deg(\Phi_2)}
\omega(\Phi_2,\Phi_1).\end{equation}
Gauge invariance requires that $n$-string products are BPZ odd:
\begin{equation}\langle \omega|\mathbb{I}\otimes L_n 
= -\langle\omega|L_n\otimes\mathbb{I}.\end{equation}
This implies that the vertices are symmetric under permutations of the 
inputs. (This is called a {\it cyclic} $L_\infty$
 algebra, though the vertices have full permutation symmetry). 
With these ingredients, we can write a gauge invariant closed string
action,
\begin{equation}S=\sum_{n=0}^\infty\frac{1}{(n+2)!}
\omega(\Phi,L_{n+1}(\underbrace{\Phi,...,\Phi}_{n+1\ \mathrm{times}})).
\end{equation}

\section{Witten's Theory with Stubs}
\label{sec:Wstubs}

In this section we revisit the construction of Witten's open superstring 
field theory. Unlike \cite{WittenSS}, where the higher vertices were built from
Witten's open string star product, here we consider a more general 
set of vertices which may include integration over bosonic moduli. Such 
vertices are at any rate necessary for the closed string \cite{noJacobi}.

Witten's superstring field theory is based on a string field $\Psi$ in the
$-1$ picture. It has even degree (but is Grassmann odd), ghost number $1$, and 
lives in the small Hilbert space. The action is defined by a sequence of 
multi-string products
\begin{equation}M_1^{(0)}\!=\!Q,\ M_2^{(1)},\ M_3^{(2)},\ M_4^{(3)},
\ ...\ ,\end{equation}
satisfying the relations of a cyclic $A_\infty$ algebra. Since the vertices 
must have total picture $-2$, and the string field has picture $-1$, the 
$(n+1)$st product $M_{n+1}^{(n)}$ must carry picture $n$.\footnote{Ghost 
number saturation is also important, but is essentially automatic in our 
construction. Suffice it to say that products $M_n^{(k)}$ carry ghost number 
$2-n$ and gauge products $\mu_n^{(k)}$ carry ghost number $1-n$ for the open 
string. For the closed string, products $L_{n}^{(p,q)}$ carry ghost number
$3-2n$ and gauge products $\lambda_{n}^{(p,q)},\lb_n^{(p,q)}$ carry ghost 
number $2-2n$.}  
We keep track of the picture through the upper index of the product. 
The goal is to construct these products by placing picture changing operators 
on a set of $n$-string products defining open bosonic string field theory:
\begin{equation}M_1^{(0)}\!=\!Q,\ M_2^{(0)},\ M_3^{(0)},\ M_4^{(0)},\ ...\ ,
\label{eq:bos_prod}\end{equation}
where the bosonic string products of course carry zero picture.
We can choose $M_2^{(0)}$ to be Witten's open string star product, in which 
case the higher bosonic products $M_3^{(0)},M_4^{(0)},...$ can be chosen to 
vanish. This is the scenario considered in \cite{WittenSS}. Here we will not 
assume that $M_3^{(0)},M_4^{(0)},...$ vanish. For example, we can consider 
the open string star product with ``stubs'' attached to each output:
\begin{equation}M_2^{(0)}(A,B) = (-1)^{\deg(A)}e^{-\pi L_0}
\Big((e^{-\pi L_0}A)*(e^{-\pi L_0}B)\Big).\end{equation}
The presence of stubs means that the propagators by themselves will not cover
the full bosonic moduli space, and the higher products 
$M_3^0,M_4^0,...$ are needed to cover the missing regions. Though it is natural
to think of the $M_n^{(0)}$s as deriving from open bosonic string field 
theory, this is not strictly necessary. We only require three formal
 properties:
\begin{description}
\item{1)} The $M_n^{(0)}$s satisfy the relations of a 
cyclic $A_\infty$ algebra.
\item{2)} The $M_n^{(0)}$s are in the small Hilbert space. 
\item{3)} The $M_n^{(0)}$s carry vanishing picture number.
\end{description}
Our task is to add picture number to the $M_n^{(0)}$s to define 
consistent nonzero vertices for Witten's open superstring field theory. 

\subsection{Cubic and Quartic Vertices}

We start with the cubic vertex, defined by a 2-product $M_2^{(1)}$ 
constructed by placing a picture changing operator $X$ once on each output of 
$M_2^{(0)}$: 
\begin{equation}M_2^{(1)}(\Psi_1,\Psi_2) \equiv 
\frac{1}{3}\Big(X M_2^{(0)}(\Psi_1,\Psi_2)+M_2^{(0)}(X \Psi_1,\Psi_2)
+M^{(0)}_2(\Psi_1,X \Psi_2)\Big).\label{eq:M2}
\end{equation}
The picture changing operator $X$ takes the following form:
\begin{equation}X\equiv \oint_{|z|=1} \frac{dz}{2\pi i}f(z)X(z),\ \ \ \ \ \ 
X(z)=Q\xi(z),\end{equation}
where $f(z)$ a 1-form which is analytic in some nondegenerate 
annulus around the unit circle, and satisfies
\begin{equation}f(z)=-\frac{1}{z^2}f\left(-\frac{1}{z}\right),\ \ \ \ \ \ \
\oint_{|z|=1} \frac{dz}{2\pi i}f(z) = 1.\end{equation}
The first relation implies that $X$ is BPZ even, and the second 
amounts to a choice of the open string coupling constant, 
which we have set to $1$. Since $Q$ and $X$ commute, $Q$ is a derivation
of $M_2^{(1)}$:
\begin{equation}[\Q,\M_2^{(1)}]=0.\end{equation}
Together with $[\Q,\Q]=0$, this means that the first two $A_\infty$ relations 
are satisfied. However, $M_2^{(1)}$ is not associative, so higher products 
$M_3^{(2)},M_4^{(3)},...$ are needed to have a consistent $A_\infty$ algebra. 

To find the higher products, the key observation is that $M_2^{(1)}$ is 
BRST exact in the large Hilbert space:\footnote{Note that the cohomology of 
$Q$ and $\eta$ is trivial in the large Hilbert space.}
\begin{equation}\M_2^{(1)} = [\Q,\mmu_2^{(1)}].\label{eq:M21}\end{equation}
Here we introduce a degree even product
\begin{equation}\mu_2^{(1)} \equiv \frac{1}{3}\Big(\xi M_2^{(0)} -M_2^{(0)}(
\xi\otimes\mathbb{I}+\mathbb{I}\otimes \xi)\Big),\end{equation}
with $\xi\equiv\oint \frac{dz}{2\pi i}f(z)\xi(z)$, which also satisfies 
\begin{equation}\M_2^{(0)} = [\n,\mmu_2^{(1)}],\end{equation}
where $\n$ is the coderivation derived from the $\eta$ zero mode. 
The fact that $M_2^{(1)}$ is BRST exact means that it can be generated by 
a gauge transformation through the space of $A_\infty$ structures 
\cite{WittenSS}. So to find a solution to the $A_\infty$ relations, all we 
have to do is complete the construction of the gauge transformation so as 
to ensure that $M_3^{(2)},M_4^{(3)},...$ are in the small Hilbert space. 
The gauge transformation is defined by $\mu_2^{(1)}$ and an 
array of higher-point products $\mu_{l}^{(k)}$ of even degree. We 
will call these ``gauge products.''\footnote{The 
notation and terminology for products used here differs from \cite{WittenSS}. 
The relation between here and there is $M_{n+1}^{(n)} = M_n$, 
$\mu_{n+2}^{(n+1)}=\Mb_{n+2}$ and $M_{n+2}^{(n)}=m_{n+2}$.}

The first nonlinear correction to the gauge transformation determines the 
3-product $M_3^{(2)}$, via the formula
\begin{equation}\M_3^{(2)} = \frac{1}{2}\Big([\Q,\mmu_3^{(2)}]
+[\M_2^{(1)},\mmu_2^{(1)}]\Big),\label{eq:M3}\end{equation}
where we introduce a gauge 3-product $\mmu_3^{(2)}$ with picture 
number two. Plugging in and using the Jacobi identity, it is 
easy to see that the 3rd $A_\infty$ relation is identically satisfied:
\begin{equation}0=\frac{1}{2}[\M_2^{(1)},\M_2^{(1)}]+[\Q,\M_3^{(2)}].
\end{equation}
However, the term $[\Q,\mmu_3^{(2)}]$ in \eq{M3} does not play a role for this
purpose. This term is needed for a different reason: to ensure that 
$M_3^{(2)}$ lives in the small Hilbert space. Let's define a degree
odd 3-product $M_3^{(1)}$ with picture 1, satisfying 
\begin{equation}\M_3^{(1)}=[\n,\mmu_3^{(2)}].\label{eq:M31}\end{equation} 
Requiring $M_3^{(2)}$ to be in the small Hilbert space implies 
\begin{eqnarray}[\n,\M_3^{(2)}]=0\lineup
=\frac{1}{2}\Big(-[\Q,\M_3^{(1)}]-[\M_2^{(1)},\M_2^{(0)}]\Big),\nonumber\\
\lineup = \frac{1}{2}\left[\Q,-\M_3^{(1)}+[\M_2^{(0)},\mmu_2^{(1)}]\right].
\label{eq:M31n}\end{eqnarray}
Therefore $M_3^{(1)}$ must satisfy
\begin{equation}\M_3^{(1)} = [\Q,\mmu_3^{(1)}]+[\M_2^{(0)},\mmu_2^{(1)}],
\end{equation}
where we introduce yet another gauge 3-product $\mu_3^{(1)}$ with 
picture number 1. In \cite{WittenSS} it was consistent to set $\mu_3^{(1)}=0$
because Witten's open string star product is associative. Now we will not 
assume that $M_2^{(0)}$ is associative, so the term $[\Q,\mmu_3^{(1)}]$ is 
needed to make sure that $M_3^{(1)}$ is in the small Hilbert space, as is 
required by \eq{M31}. We define $\mu_3^{(1)}$ by the relation
\begin{equation}2\M_3^{(0)}=[\n,\mmu_3^{(1)}],\label{eq:M30n}\end{equation}
where $\M_3^{(0)}$ is the bosonic 3-product. Then taking $\eta$ of \eq{M31n}
implies
\begin{equation}0=[\Q,\M_3^{(0)}]+\frac{1}{2}[\M_2^{(0)},\M_2^{(0)}].
\end{equation}
This is nothing but the 3rd $A_\infty$ relation for the bosonic products. The
upshot is that we can determine $M_3^{(2)}$ for Witten's superstring
field theory by climbing a ``ladder'' of products and gauge products 
starting from  $M_3^{(0)}$ as follows:
\begin{eqnarray}
\M_3^{(0)}\lineup = \mathrm{given},\\
\mu_3^{(1)} \lineup = \frac{1}{2}\Big(\xi M_3^{(0)}-M_3^{(0)}
(\xi\otimes\mathbb{I}\otimes\mathbb{I}
+\mathbb{I}\otimes\xi\otimes\mathbb{I}
+\mathbb{I}\otimes\mathbb{I}\otimes\xi)\Big),\\
\M_3^{(1)}\lineup = [\Q,\mmu_3^{(1)}]+[\M_2^{(0)},\mmu_2^{(1)}],
\label{eq:M3ex}\\
\mu_3^{(2)} \lineup = \frac{1}{4}\Big(\xi M_3^{(1)}-M_3^{(1)}
(\xi\otimes\mathbb{I}\otimes\mathbb{I}
+\mathbb{I}\otimes\xi\otimes\mathbb{I}
+\mathbb{I}\otimes\mathbb{I}\otimes\xi)\Big),\\
\M_3^{(2)} \lineup = \frac{1}{2}\Big([\Q,\mmu_3^{(2)}]
+[\M_2^{(1)},\mmu_2^{(1)}]\Big).\label{eq:M31ex}
\end{eqnarray}
The second and fourth equations invert \eq{M30n} and \eq{M31} 
by placing a $\xi$ insertion once on each output of the respective 
3-product. Incidentally, we construct $M_2^{(1)}$ by climbing a similar 
ladder
\begin{eqnarray}
\M_2^{(0)}\lineup = \mathrm{given},\\
\mu_2^{(1)} \lineup 
= \frac{1}{3}\Big(\xi M_2^{(0)} -M_2^{(0)}(
\xi\otimes\mathbb{I}+\mathbb{I}\otimes \xi)\Big),\\
\M_2^{(1)}\lineup = [\Q,\mmu_2^{(1)}],
\end{eqnarray}
but in this case it was easier to postulate the final answer from the 
beginning, \eq{M2}.

\begin{figure}
\begin{center}
\resizebox{2.9in}{3in}{\includegraphics{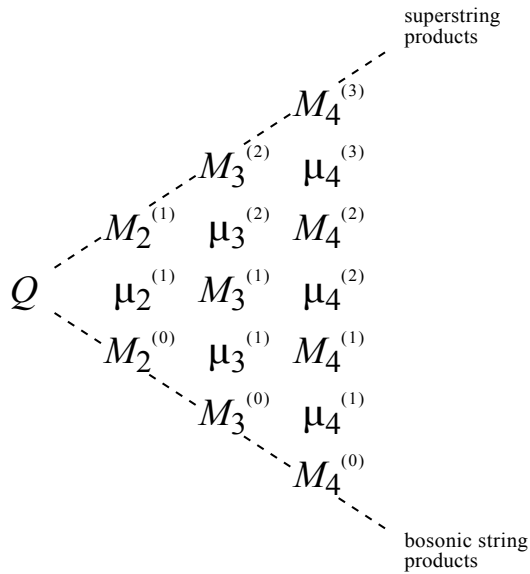}}
\end{center}
\caption{\label{fig:triangle} The products $M_n^{(n-1)}$ 
of Witten's superstring field theory are derived from the 
products  $M_n^{(0)}$ of open bosonic string field theory by constructing 
a triangular array of products of intermediate picture number.}
\end{figure}

Proceeding in this way, it is not difficult to anticipate that the 
$(n+1)$-string product $M_{n+1}^{(n)}$ of Witten's superstring field theory 
can be constructed by ascending a ladder of $n+1$ products
\begin{equation}M_{n+1}^{(0)},\ M_{n+1}^{(1)},\ ... ,\ M_{n+1}^{(n)},
\end{equation}
interspersed with $n$ gauge products 
\begin{equation}\mu_{n+1}^{(1)},\ \mu_{n+1}^{(2)},\ ...,\ \mu_{n+1}^{(n)},
\end{equation}
adding picture number one step at a time. Thus we will have 
a recursive solution to the $A_\infty$ relations, expressed in terms of 
a ``triangle'' of products, as shown in figure \ref{fig:triangle}.

\subsection{All Vertices}
\label{subsec:All}

We now explain how to determine the vertices to to all orders. 
We start by collecting superstring products into a generating function
\begin{equation}\M^{[0]}(t)\equiv \sum_{n=0}^\infty t^n\M_{n+1}^{(n)},
\label{eq:M0}\end{equation}
so that the $(n+1)$st superstring product can be extracted by looking at the 
coefficient of $t^n$. Here we place an upper index on the generating function
(in square brackets) to indicate the ``deficit'' in picture number of 
the products relative to what is needed for the superstring. 
In this case, of course, the deficit is zero. The superstring products must 
satisfy two properties. First, they must be in the small Hilbert
space, and second, they must satisfy the $A_\infty$ relations:
\begin{equation}[\n,\M^{[0]}(t)]=0,\ \ \ \ \ [\M^{[0]}(t),\M^{[0]}(t)]=0.
\label{eq:supsmalAinf}\end{equation}
Expanding the second equation in powers of $t$ gives the $A_\infty$ relations 
as written in \eq{Ainf}. To solve the $A_\infty$ relations, we postulate 
the differential equation
\begin{equation}\frac{\d}{\d t}\M^{[0]}(t) = [\M^{[0]}(t),\mmu^{[0]}(t)],
\label{eq:M0dt}\end{equation}
where 
\begin{equation}\mmu^{[0]}(t) = \sum_{n=0}^\infty t^n\mmu_{n+2}^{(n+1)}
\end{equation}
is a generating function for ``deficit-free'' gauge products. Expanding 
\eq{M0dt} in powers of $t$ gives previous formulas \eq{M21} and \eq{M31} 
for the 2-product and the 3-product. Note that this differential equation 
implies
\begin{equation}\frac{\d}{\d t}[\M^{[0]}(t),\M^{[0]}(t)] 
= [[\M^{[0]}(t),\M^{[0]}(t)],\mmu^{[0]}(t)].
\end{equation}
Since this is homogeneous in $[\M^{[0]}(t),\M^{[0]}(t)]$, the $A_\infty$ 
relations follow immediately from the fact that $[\M^{[0]}(t),\M^{[0]}(t)]=0$ 
holds at $t= 0$ (since $\Q$ is nilpotent). Note that the generating 
function \eq{M0} can also be interpreted as defining a 1-parameter family of 
$A_\infty$ algebras, where the parameter $t$ is the open string coupling 
constant \cite{WittenSS}. In this context, the differential equation 
\eq{M0dt} says that changes of the coupling constant are implemented by a 
gauge transformation through the space of $A_\infty$ structures, and 
$\mmu^{[0]}(t)$ is the infinitesimal gauge parameter. 

The statement that the coupling constant is ``pure gauge'' normally means 
that the cubic and higher order vertices can be removed by field redefinition,
and the scattering amplitudes vanish \cite{Linfgauge}. This 
does not happen here because $\mmu^{[0]}(t)$ is in the large Hilbert space, 
and therefore does not define an ``admissible'' gauge parameter. But 
then the nontrivial condition is that the superstring 
products are in the small Hilbert space despite the fact that the gauge 
transformation defining them is not. To see what this condition implies, 
take $\n$ of the differential equation \eq{M0dt} to find
\begin{equation}[\M^{[0]}(t),\M^{[1]}(t)]=0,\label{eq:M0M1}\end{equation}
where 
\begin{equation}\M^{[1]}(t)=[\n,\mmu^{[0]}(t)]
=\sum_{n=0}^\infty t^n \M_{n+2}^{(n)}\end{equation}
is the generating function for products with a single picture deficit. 
Now we can solve \eq{M0M1} by postulating a new differential equation
\begin{equation}\frac{\d}{\d t}\M^{[1]}(t) = [\M^{[0]}(t),\mmu^{[1]}(t)]+
 [\M^{[1]}(t),\mmu^{[0]}(t)],
\label{eq:M1dt}\end{equation}
where 
\begin{equation}\mmu^{[1]}(t)=\sum_{n=0}^\infty t^n\mmu_{n+3}^{(n+1)}
\end{equation}
is a generating function for gauge products with a single picture deficit. 
Now we are beginning to see the outlines of a recursion. Taking $\n$ of 
\eq{M1dt} implies a constraint on the generating function for 
products with two picture deficits $\M^{[2]}(t)$, which can be solved by 
postulating yet another differential equation, and so on. The full recursion 
is most compactly expressed by packaging the generating functions $\M^{[m]}(t)$
and $\mmu^{[m]}(t)$ together in a power series in a new parameter $s$:
\begin{eqnarray}\M(s,t) \lineup \equiv \sum_{m=0}^\infty s^m\M^{[m]}(t)=
\sum_{m,n=0}^\infty s^m t^n\M^{(n)}_{m+n+1},\\
\mmu(s,t)\lineup \equiv \sum_{m=0}^\infty s^m\mmu^{[m]}(t)
=\sum_{m,n=0}^\infty s^m t^n\mmu_{m+n+2}^{(n+1)}.
\end{eqnarray}
Note that powers of $t$ count the picture number, and powers of $s$ count 
the deficit in picture number. At $t=0$ $\M(s,t)$ reduces to a generating 
function for products of the bosonic string, and at $s=0$ it reduces to 
a generating function for products of the superstring:
\begin{eqnarray}\M(s,0) \lineup = \sum_{n=0}^\infty s^n\M_{n+1}^{(0)},\\
\M(0,t)\lineup = \M^{[0]}(t) = \sum_{n=0}^\infty t^n\M_{n+1}^{(n)}.
\end{eqnarray}
The recursion then emerges from expansion of a {\it pair} of 
differential equations 
\begin{eqnarray}\frac{\d}{\d t}\M(s,t)\lineup 
=[\M(s,t),\mmu(s,t)],\label{eq:Mdt}\\
\frac{\d}{\d s}\M(s,t)\lineup = [\n,\mmu(s,t)].\label{eq:Mds}
\end{eqnarray}
Note that these equations imply 
\begin{eqnarray}
\frac{\d}{\d t}[\M(s,t),\M(s,t)]\lineup 
=[[\M(s,t),\M(s,t)],\mmu(s,t)],\\
\frac{\d}{\d t}[\n,\M(s,t)]\lineup = [[\n,\M(s,t)],\mmu(s,t)]-\frac{1}{2}
\frac{\d}{\d s}[\M(s,t),\M(s,t)].\label{eq:dnm}
\end{eqnarray}
Since the first equation is homogeneous in $[\M(s,t),\M(s,t)]$, the 
$A_\infty$ relations for the bosonic products at $t=0$ implies
$[\M(s,t),\M(s,t)]=0$ for all $s$ and $t$. Thus the second equation 
\eq{dnm} becomes homogeneous in $[\n,\M(s,t)]$, and the fact 
that the bosonic products are in the small Hilbert space at $t=0$ implies 
that all products are in the small Hilbert space. Thus 
\begin{equation}[\M(s,t),\M(s,t)]=0,\ \ \ \ [\n,\M(s,t)]=0.\end{equation}
Setting $s=0$ we recover \eq{supsmalAinf}. Therefore, solving \eq{Mdt} 
and \eq{Mds} automatically determines a set of 
superstring products which live in the small Hilbert 
space and satisfy the $A_\infty$ relations.  

Now all we need to do is solve the differential equations 
\eq{Mdt} and \eq{Mds} to determine the products. Expanding \eq{Mdt} in 
$s,t$ and reading off the coefficient of $s^m t^n$ gives the formula:
\begin{equation}\M_{m+n+2}^{(n+1)}=\frac{1}{n+1}\sum_{k=0}^n
\sum_{l=0}^m[\M_{k+l+1}^{(k)},\mmu_{m+n-k-l+2}^{(n-k+1)}].\label{eq:Mrec}
\end{equation}
This determines the product $M_{m+n+2}^{(n+1)}$ if we are given gauge 
products
\begin{equation}\mu_{l}^{(k)},\ \ \ \ \ 1\leq k\leq n+1,
\ \ \ k+1\leq l\leq k+m+1,\end{equation}
and the lower order products 
\begin{equation}
 M_{l}^{(k)},\ \ \ \ \ 0\leq k\leq n,\ \ \ k+1\leq l\leq k+m+1.
\end{equation}
The lower order products are either again determined by \eq{Mrec},
or they are products of the bosonic string, which we assume are given. So 
now we must find the gauge products $\mu_l^{(k)}$. Expanding \eq{Mds}
gives
\begin{equation}
[\n,\mmu_{m+n+2}^{(n+1)}] = (m+1)\M_{m+n+2}^{(n)}.\label{eq:Mbrec}
\end{equation}
This equation will determine $\mu_{m+n+2}^{(n+1)}$ in terms of 
$M_{m+n+2}^{(m)}$. The solution is not unique. However there is 
a natural ansatz preserving cyclicity: 
\begin{equation}\mu_{m+n+2}^{(m+1)} = \frac{n+1}{m+n+3}
\left(\xi M_{m+n+2}^{(m)}- M_{m+n+2}^{(m)}\sum_{k=0}^{m+n+1}
\mathbb{I}^{\otimes k}\otimes \xi\otimes\mathbb{I}^{\otimes m+n+1-k}\right).
\label{eq:Mb}\end{equation}
or, more compactly, we can write 
$\mmu_{m+n+2}^{(m+1)}=(n+1)\xi\circ\M_{m+n+2}^{(m)}$
where $\xi\circ$ denotes the operation of taking the average of $\xi$ 
acting on the output and on each input of the product.
This ansatz works assuming $M_{m+n+2}^{(m)}$ is in the small Hilbert space, 
but we have to show that the ansatz is consistent with that assumption. 
To this end, note that if \eq{Mdt} is satisfied and the gauge products are 
defined by \eq{Mb}, we have the relation  
\begin{equation}
\frac{\d}{\d t}[\n,\M(s,t)]  = [[\n,\M(s,t)],\mmu(s,t)]+\left[\M(s,t),
\frac{\d}{\d s}\xi\circ[\n,\M(s,t)]\right].
\end{equation}
Since this equation is homogeneous in $[\n,\M(s,t)]$, \eq{Mb} implies that 
all products must be in the small Hilbert space.

The construction is recursive. Assume that we have already
constructed all products $M_{m}^{(k)}$ and gauge products 
$\mu_{m}^{(k)}$ with $m\leq n$ inputs and with all picture 
numbers. Then we construct the $(n+1)$st product of Witten's superstring 
field theory by climbing a ladder of products and gauge products, defined 
by equations \eq{Mrec} and \eq{Mbrec}:
\begin{eqnarray}
\M_{n+1}^{(0)}\lineup = \mathrm{given},\nonumber\\
\mu_{n+1}^{(1)}\lineup = \frac{n}{n+2}
\left(\xi M_{n+1}^{(0)}- M_{n+1}^{(0)}\sum_{k=0}^{n}
\mathbb{I}^{\otimes k}\otimes \xi\otimes\mathbb{I}^{\otimes n-k}\right)
.\nonumber\\
\M_{n+1}^{(1)}\lineup = [\Q,\mmu_{n+1}^{(1)}]+[\M_2^{(0)},\mmu_{n}^{(1)}]
+...+[\M_{n}^{(0)},\mmu_2^{(1)}],\nonumber\\
\mu_{n+1}^{(2)}\lineup = \frac{n-1}{n+2}
\left(\xi M_{n+1}^{(1)}- M_{n+1}^{(1)}\sum_{k=0}^{n}
\mathbb{I}^{\otimes k}\otimes \xi\otimes\mathbb{I}^{\otimes n-k}\right),
\nonumber\\
\M_{n+1}^{(2)}\lineup =\frac{1}{2}\Big([\Q,\mmu_{n+1}^{(2)}]
+[\M_2^{(0)},\mmu_{n}^{(2)}]+[\M_2^{(1)},\mmu_{n}^{(1)}]+...\nonumber\\
\lineup\ \ \ \ \ \ \ \ +[\M_{n-1}^{(0)},\mmu_3^{(2)}]
+[\M_{n-1}^{(1)},\mmu_3^{(1)}]+[\M_{n}^{(1)},\mmu_2^{(1)}]\Big),\nonumber\\
\lineup \vdots\nonumber\\
\mu_{n+1}^{(n)}\lineup = \frac{1}{n+2}
\left(\xi M_{n+1}^{(n)}- M_{n+1}^{(n)}\sum_{k=0}^{n}
\mathbb{I}^{\otimes k}\otimes \xi\otimes\mathbb{I}^{\otimes n-k}\right),
\nonumber\\
\M_{n+1}^{(n)} \lineup 
= \frac{1}{n}\Big([\Q,\mmu_{n+1}^{(n)}]+[\M_2^{(1)},
\mmu_{n}^{(n-1)}]+...+[\M_{n}^{(n-1)},\mmu_{2}^{(1)}]\Big).
\end{eqnarray}
The final step in this ladder is the $n+1$-string product of Witten's open 
superstring field theory. Incidentally, note that the nature of this 
construction guarantees that the superstring products will define cyclic 
vertices if the bosonic products do (see appendix B of 
\cite{WittenSS}).

\section{NS Heterotic String}
\label{sec:het}

Our analysis of the open superstring almost immediately generalizes to a 
construction of heterotic string field theory in the NS sector. An 
alternative formulation of this theory, using the large Hilbert space, is 
described in \cite{heterotic1,heterotic}.
The closed string field is a degree (and Grassmann) even NS state $\Phi$ 
in the superconformal field theory of a heterotic string. Note that the
$\beta\gamma$ ghosts and picture only reside in the leftmoving sector. 
The string field has ghost number $2$ and picture number $-1$, and 
satisfies the $b_0^-$ and level matching constraints \eq{level}. An 
on-shell state in Siegel gauge takes the form
\begin{equation}\Phi\sim c\overline{c}e^{-\phi}\mathcal{O}^{\mathrm{m}}(0,0),
\end{equation}
where $\mathcal{O}^\mathrm{m}$ is a matter primary operator with 
left/rightmoving dimension $(\frac{1}{2},1)$. The symplectic form 
\eq{symp_cl} is nonvanishing only on states whose ghost number adds 
up to five and whose picture number adds up to $-2$. 

The action is defined by a sequence of degree odd closed string products
\begin{equation}L_1^{(0)}\!=\!Q,\ \ L_2^{(1)},\ \ L_3^{(2)},\ \ L_4^{(3)},\ \ 
...\ ,\end{equation}
satisfying the relations of a cyclic $L_\infty$ algebra. Just like in the 
open string, the $n$th closed string product must have picture $n-1$ to 
define a nonvanishing vertex. We construct the products by placing
picture changing operators on the products of the closed bosonic string 
\begin{equation}L_1^{(0)}\!=\!Q,\ \ L_2^{(0)},\ \ L_3^{(0)},\ \ L_4^{(0)},\ \ 
...\ ,\end{equation}
which, or course, have vanishing picture. The explicit definition of 
the closed bosonic string products is an intricate story 
\cite{ZwiebachClosed,Saadi,MoellerQuartic,MoellerQuinticI,MoellerQuinticII}, 
but for our purposes all we need to know is:
1) they satisfy the relations of a cyclic $L_\infty$ algebra,
2) they are in the small Hilbert space, 3) they 
carry vanishing picture number, and 4) they are consistent with the 
$b_0^-$ and $L_0^-$ constraints. 

The problem we need to solve appears completely analogous to the open 
superstring. Aside from replacing tensor products with wedge 
products, there is one minor difference. Since the products of the heterotic 
string must respect the $b_0^-$ and $L_0^-$ constraints, the picture changing
operator $X$ in the 2-product
\begin{equation}L_2^{(1)}(\Phi_1,\Phi_2) = 
\frac{1}{3}\Big(XL_2^{(0)}(\Phi_1,\Phi_2)+L_2^{(0)}(X\Phi_1,\Phi_2)
+L_2^{(0)}(\Phi_1,X\Phi_2)\Big)\end{equation}
must be identified with the zero mode $X_0$. This way, we can pull 
$b_0^-$ and $L_0^-$ past $X_0$ to act on $L_2^{(0,0)}$, which vanishes. 
More generally, we must construct closed superstring products using the $\xi$ 
zero mode 
\begin{equation}\xi=\xi_0 = \oint_{|z|=1}\frac{dz}{2\pi i}\frac{1}{z}\xi(z),
\end{equation}
rather than a more general charge which would be consistent for the open 
string.

Following the discussion of the open superstring, we introduce a ``triangle''
of products
\begin{equation}L_{n+1}^{(k)},\ \ \ \  0\leq n\leq\infty,\ \ \ 0\leq k\leq n,
\end{equation}
and gauge products,
\begin{equation}\lambda_{n+2}^{(k+1)},\ \ \ \  
0\leq n\leq\infty,\ \ \ 0\leq k\leq n
\end{equation}
of intermediate picture indicated in the upper index. We build the 
$(n+1)$-heterotic string product $L_{n+1}^{(n)}$ by climbing a ``ladder''
of products 
\begin{equation}L_{n+1}^{(0)},\ \lambda_{n+1}^{(1)},\ L_{n+1}^{(1)},\ ...,\
\lambda_{n+1}^{(n)},\ L_{n+1}^{(n)},\end{equation}
adding picture one step at a time. Each step is prescribed by the 
closed string analogues of equations \eq{Mrec} and \eq{Mbrec}:
\begin{eqnarray}\L_{m+n+2}^{(m+1)}\lineup =\frac{1}{m+1}\sum_{k=0}^m
\sum_{l=0}^n[\L_{k+l+1}^{(k)},\ll_{m+n-k-l+2}^{(m-k+1)}]\label{eq:Lrec}\\
\lambda_{m+n+2}^{(m+1)} \lineup = \frac{n+1}{m+n+3}
\Big(\xi_0 L_{m+n+2}^{(m)}- L_{m+n+2}^{(m)}(\xi_0\w\mathbb{I}_{m+n+1})\Big).
\label{eq:Lbrec}
\end{eqnarray}
The only differences from the open superstring are that the coderivations 
act on the symmetrized tensor algebra, and $\xi$ has been replaced by $\xi_0$.

\section{NS-NS Closed Superstring}
\label{sec:NSNS}

We are now ready to discuss the NS-NS sector of Type II closed superstring 
field theory. A recent proposal for defining this theory in the large Hilbert
space appears in \cite{Matsunaga}. The closed string field is a degree 
even (and Grassmann even) NS-NS state $\Phi$ in the superconformal field 
theory of a type II superstring.
Now $\beta\gamma$ ghosts and picture occupy both the leftmoving and rightmoving
sectors. The string field has ghost number $2$, satisfies the $b_0^-$ and 
$L_0^-$ constraints \eq{level}, and has left/rightmoving 
picture number $(-1,-1)$. On-shell states in Siegel gauge 
take the form
\begin{equation}\Phi\sim c\overline{c}e^{-\phi}e^{-\overline{\phi}}
\mathcal{O}^\mathrm{m}(0,0),\end{equation}
where $\mathcal{O}^\mathrm{m}$ is a superconformal matter primary of weight
$(\frac{1}{2},\frac{1}{2})$. The symplectic form \eq{symp_cl} is nonvanishing
on states of ghost number $5$ and left/right picture $(-2,-2)$.

The theory is defined by a sequence of degree odd closed string products
\begin{equation}L_1^{(0,0)}\!=\!Q,\ \ L_2^{(1,1)},\ \ L_3^{(2,2)},
\ \ L_4^{(3,3)},\ \ ...\ ,\end{equation}
satisfying the relations of a cyclic $L_\infty$ algebra. The $(n+1)$st closed 
string product must have left/right picture $(n,n)$. These products should 
be constructed from the products of the closed bosonic string,
\begin{equation}L_1^{(0,0)}\!=\!Q,\ \ L_2^{(0,0)},\ \ L_3^{(0,0)},\ \ 
L_4^{(0,0)},\ \ ...\ ,\end{equation}
which have vanishing picture. Note that we add an extra index to indicate 
rightmoving picture. Now the situation is somewhat different from the 
open string, since we need to add twice as much picture and we need to pay 
attention to how it is distributed between leftmoving and rightmoving
sectors. However, it is not difficult to guess what the 2-product should 
look like. Starting with $L_2^{(0,0)}$, we surround it once with a leftmoving 
picture changing operator $X_0$, and again a rightmoving picture changing 
operator $\overline{X}_0$, to produce the expression
\begin{eqnarray}
L_2^{(1,1)}(\Phi_1,\Phi_2)\lineup =\frac{1}{9}\Big(
X_0\overline{X}_0L_2^{(0,0)}(\Phi_1,\Phi_2)
+X_0L_2^{(0,0)}(\overline{X}_0\Phi_1,\Phi_2)
+X_0L_2^{(0,0)}(\Phi_1,\overline{X}_0\Phi_2)\nonumber\\
\lineup\ \ \ \ \ +\overline{X_0}L_2^{(0,0)}(X_0\Phi_1,\Phi_2)
+L_2^{(0,0)}(X_0\overline{X}_0\Phi_1,\Phi_2)
+L_2^{(0,0)}(X_0\Phi_1,\overline{X}_0\Phi_2)\nonumber\\
\lineup\ \ \ \ \ +\overline{X_0}L_2^{(0,0)}(\Phi_1,X_0\Phi_2)
+L_2^{(0,0)}(\overline{X}_0\Phi_1,X_0\Phi_2)
+L_2^{(0,0)}(\Phi_1,X_0\overline{X}_0\Phi_2)\Big).\nonumber\\
\label{eq:L211}\end{eqnarray}
Note that since $X_0$ and $\overline{X}_0$ commute it does not matter which 
order we apply them to the bosonic product.  

\subsection{Asymmetric Construction}
\label{subsec:comp}

The easiest solution for the closed superstring is to apply the open 
string construction twice: The first time to 
get the correct picture number for leftmovers and a second time to get the 
correct picture number for the rightmovers. More specifically we proceed as 
follows. Starting with the bosonic product $L_{n+1}^{(0,0)}$ we climb 
a ``ladder'' of products and gauge products
\begin{equation}L_{n+1}^{(0,0)},\ \lambda_{n+1}^{(1,0)},\ L_{n+1}^{(1,0)},
\ ..., \ \lambda_{n+1}^{(n,0)},L_{n+1}^{(n,0)},
\end{equation}
using \eq{Lrec} and \eq{Lbrec} (with an extra spectator index for rightmoving
picture). At the top of the ladder, the product $L_{n+1}^{(n,0)}$ has the 
required leftmoving picture, but the rightmoving picture is still absent. 
So we take $L_{n+1}^{(n,0)}$ as the input for a second set of recursions 
which add rightmoving picture. Starting with $L_{n+1}^{(n,0)}$ we climb 
a second ``ladder''
\begin{equation}L_{n+1}^{(n,0)},
\ \lambda_{n+1}^{(n,1)},\ L_{n+1}^{(n,1)},\ ...,
\ \lambda_{n+1}^{(n,n)},L_{n+1}^{(n,n)},
\end{equation}
again using \eq{Lrec} and \eq{Lbrec}, but this time the leftmoving 
picture is a spectator index, and the right moving zero mode $\overline{\xi}_0$
appears in \eq{Lbrec} rather than the leftmoving one. Thus, for example 
the 2-product is constructed by climbing two ladders:
\begin{eqnarray}
\mathrm{first\ ladder}\lineup  
\left\{\begin{array}{l}
\L_2^{(0,0)}=\mathrm{given},\\[8pt]
\lambda_2^{(1,0)} = \displaystyle{\frac{1}{3}}\Big(\xi_0 L_2^{(0,0)}
-L_2^{(0,0)}(\xi_0\w\mathbb{I})\Big),\\[8pt]
\L_2^{(1,0)}=[\Q,\ll_2^{(1,0)}],
\end{array}\right.
\nonumber\\
\mathrm{second\ ladder}\lineup  
\left\{\begin{array}{l}
\L_2^{(1,0)}=\mathrm{given\ by\ first\ ladder}, \\ [8pt]
\lambda_2^{(1,1)} = \displaystyle{\frac{1}{3}}\Big(\overline{\xi}_0 L_2^{(1,0)}
-L_2^{(1,0)}(\overline{\xi}_0\w\mathbb{I})\Big),\\ [8pt]
\L_2^{(1,1)}=[\Q,\ll_2^{(1,1)}].
\end{array}
\right.
\end{eqnarray}
This is the simplest construction we have found the NS-NS superstring, in 
the sense that it requires the fewest auxiliary products of intermediate 
picture number in defining the recursion. However, it suffers from a curious 
asymmetry between left and rightmoving picture changing operators. This 
asymmetry first appears in $L_3^{(2,2)}$, which for example has a term of 
the form
\begin{equation}
\overline{X}_0^2L_2^{(0,0)}\Big(X_0\xi_0L_2^{(0,0)}\Big(\Phi_1,\Phi_2\Big)
,\Phi_3\Big),
\end{equation}
and no corresponding term with left and rightmovers reversed. 

\subsection{Symmetric Construction}

\begin{figure}
\begin{center}
\resizebox{6in}{2.6in}{\includegraphics{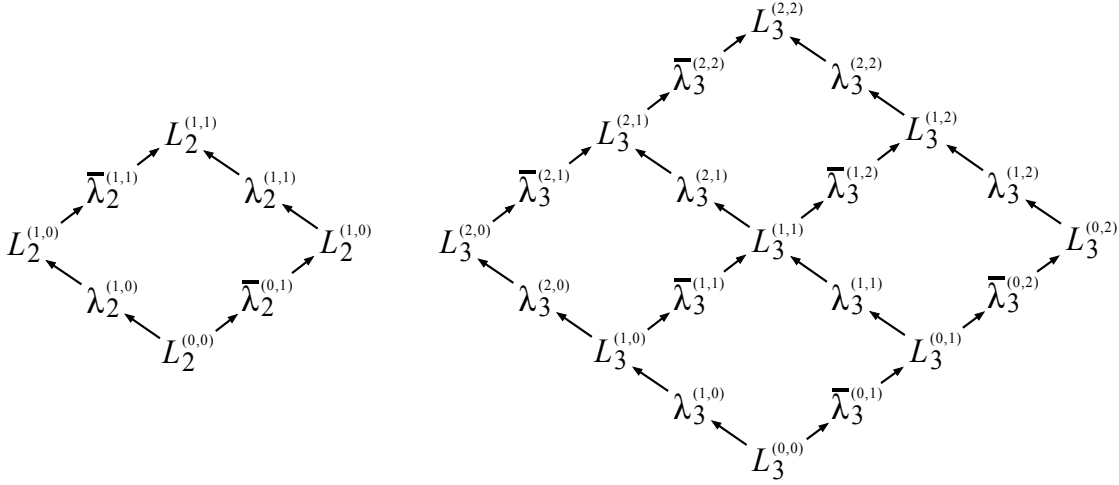}}
\end{center}
\caption{\label{fig:diamond} ``Diamonds'' of products and gauge products 
needed to construct the 2-product and 3-product of NS-NS closed 
superstring field theory.}
\end{figure}

To restore symmetry between left and rightmovers we consider a different 
solution of the $L_\infty$ relations. To motivate the structure, 
consider the 2-product $L_2^{(1,1)}$ written in the form
\begin{equation}\L_2^{(1,1)}
=\frac{1}{2}[\Q,\ll_2^{(1,1)}+\llb_2^{(1,1)}].\end{equation}
Now we have introduced {\it two} gauge products. The first $\l_2^{(1,1)}$ will
be called a ``left'' gauge product, and is defined by replacing $X_0$ in 
the expression \eq{L211} for $L_2^{(1,1)}$ with $\xi_0$. The second 
$\lb_2^{(1,1)}$ will be called a ``right'' gauge product, and is defined by
replacing $\overline{X}_0$ in $L_2^{(1,1)}$ with $\overline{\xi}_0$. 
Once we act with $\Q$, $\l_2^{(1,1)}$ and $\lb_2^{(1,1)}$ produce the same 
expression (hence the factor of $1/2$), but the advantage of this 
decomposition is that left/right symmetry is manifest. Denoting the 
left/rightmoving eta zero modes by $\eta$ and $\overline{\eta}$, we have 
the relations
\begin{eqnarray}\lineup[\n,\ll_2^{(1,1)}] = \L_2^{(0,1)}
\ \ \ \ \ [\nb,\llb_2^{(1,1)}]=\L_2^{(1,0)}\\
\lineup[\n,\llb_2^{(1,1)}] = 0
\ \ \ \ \ \ \ \ \ \ [\nb,\ll_2^{(1,1)}]=0.\end{eqnarray}
Note that the left gauge product $\l_2^{(1,1)}$ is in the 
``rightmoving small Hilbert space,'' while the right gauge product
$\lb_2^{(1,1)}$ is in the ``leftmoving small Hilbert space.'' The 
products $L_2^{(1,0)}$ and $L_2^{(0,1)}$ now carry a single 
$X_0$ or $\overline{X}_0$ insertion, respectively. Pulling $\Q$
out we can write
\begin{equation}\L_2^{(1,0)} = [\Q,\ll_2^{(1,0)}],\ \ \ \ \ 
\L_2^{(0,1)}=[\Q,\llb_2^{(0,1)}],\end{equation}
where $\l_2^{(1,0)}$ and $\lb_2^{(0,1)}$ are left/right gauge products 
satisfying 
\begin{eqnarray}\lineup [\n,\ll_2^{(1,0)}]=[\nb,\llb_2^{(0,1)}]=\L_2^{(0,0)}\\
\ \lineup [\nb,\ll_2^{(1,0)}]=[\n,\llb_2^{(0,1)}]=0,
\end{eqnarray}
and $L_2^{(0,0)}$ is the product of the bosonic string. In this 
way the superstring product $L_2^{(1,1)}$ is derived by filling a 
``diamond'' of products and gauge products, as 
shown in figure \ref{fig:diamond}. 

Also shown is a ``diamond'' illustrating the derivation of the 3-product, which
has four ``cells'' giving a total of 21 intermediate products. The explicit 
formulas associated with this diagram are difficult to guess, so we will 
proceed to motivate the general construction. To find the closed superstring 
product $L_{n+1}^{(n,n)}$, we need a diamond consisting of $(n+1)^2$ products
\begin{equation}L_{n+1}^{(p,q)},\ \ \ \ 0\leq p,q\leq n,\end{equation}
$n(n+1)$ left gauge products
\begin{equation}\l_{n+1}^{(p,q)},\ \ \ \ \begin{matrix}1\leq p \leq n, \\
0\leq q\leq n,\end{matrix}\end{equation}
and $n(n+1)$ right gauge products
\begin{equation}\lb_{n+1}^{(p,q)},\ \ \ \ \begin{matrix}0\leq p \leq n, \\
1\leq q\leq n.\end{matrix}\end{equation}
We would like to package the products into three generating functions
\begin{equation}\L(s,\overline{s},t),\ \ \ \ll(s,\overline{s},t),\ \ \ 
\llb(s,\overline{s},t),\end{equation}
which depend on three variables, corresponding to the three indices 
characterizing the products. The variable $t$ counts the total 
picture number, $s$ the deficit in leftmoving picture number, and 
$\overline{s}$ the deficit in rightmoving picture number. Thus
we have
\begin{eqnarray}\L(s,\overline{s},t)\lineup 
=\sum_{N=0}^\infty \sum_{i,j=0}^N t^{i+j}
s^{N-i}\,\overline{s}^{N-j}\L_{N+1}^{(i,j)},\\
\ll(s,\overline{s},t)\lineup 
=\sum_{N=0}^\infty \sum_{i=0}^N \sum_{j=0}^{N+1}t^{i+j}
s^{N-i}\,\overline{s}^{N+1-j}\ll_{N+2}^{(i+1,j)},\\
\llb(s,\overline{s},t)\lineup 
=\sum_{N=0}^\infty \sum_{i=0}^{N+1}\sum_{j=0}^N t^{i+j}
s^{N+1-i}\,\overline{s}^{N-j}\llb_{N+2}^{(i,j+1)}.
\end{eqnarray}
The ranges of summation here are complicated in comparison to 
what appears in the generating functions of the open string. The reason 
is that the closed superstring has left/rightmoving sectors with separate 
picture numbers, but not separate notions of multiplication. However, we 
can simplify these formulas by formally introducing an extra index to indicate 
``rightmoving multiplication:''
\begin{eqnarray}
\L_{m+1,n+1}^{(p,q)}\lineup \equiv \delta_{m,n}\,\L_{m+1}^{(p,q)},\\
\ll_{m+2,n+1}^{(p,q)}\lineup \equiv \delta_{m+1,n}\,\ll_{m+2}^{(p,q)},\\
\llb_{m+1,n+2}^{(p,q)}\lineup \equiv \delta_{m,n+1}\,\llb_{n+2}^{(p,q)},
\end{eqnarray}
with a Kronecker delta to identify multiplication between the left and right.
Then the generating functions take the form:
\begin{eqnarray}
\L(t,s,\overline{s})\lineup =\sum_{m,n=0}^\infty\sum_{p,q=0}^\infty 
\Big(t^m s^n\Big)
\Big(t^p\,\overline{s}^q\Big)\L_{m+n+1,p+q+1}^{(m,p)},\\
\ll(t,s,\overline{s})\lineup =\sum_{m,n=0}^\infty\sum_{p,q=0}^\infty 
\Big(t^ms^n\Big)
\Big(t^p\,\overline{s}^q\Big)\ll_{m+n+2,p+q+1}^{(m+1,p)},\\
\llb(t,s,\overline{s})\lineup =\sum_{m,n=0}^\infty\sum_{p,q=0}^\infty 
\Big(t^ms^n\Big)
\Big(t^p\,\overline{s}^q\Big)\llb_{m+n+1,p+q+2}^{(m,p+1)}.\label{eq:NSNSgen}
\end{eqnarray}
The solution to the $L_\infty$ relations is defined by the system of equations
\begin{eqnarray}
\frac{\d}{\d t}\L(s,\overline{s},t)\lineup 
=\Big[\L(s,\overline{s},t),\ll(s,\overline{s},t)+\llb(s,\overline{s},t)\Big],
\label{eq:Ldt} \\
\frac{\d}{\d s}\L(s,\overline{s},t)\lineup = [\n,\ll(s,\overline{s},t)],\ \ 
\ \ \ [\nb,\ll(s,\overline{s},t)]=0,\label{eq:Lds}\\
\frac{\d}{\d \overline{s}}\L(s,\overline{s},t) \lineup =
[\nb,\llb(s,\overline{s},t)],\ \ \ \ \ [\n,\llb(s,\overline{s},t)]=0.
\label{eq:Ldbs}
\end{eqnarray}
Note that $\L(s,\overline{s},t)$ at $t=0$ reduces to a generating 
function for bosonic products:
\begin{equation}\L(s,\bar{s},0) 
= \sum_{n=0}^\infty (s\overline{s})^n\L_{n+1}^{(0,0)}.\end{equation}
Following the argument given in section \ref{subsec:All}, this boundary 
condition together with the differential equations \eq{Ldt}-\eq{Ldbs} imply
\begin{equation}[\L(s,\overline{s},t),\L(s,\overline{s},t)]=0,\ \ \ \ \ \ 
[\n,\L(s,\overline{s},t)]=0,\ \ \ \ \ \ [\nb,\L(s,\overline{s},t)]=0. 
\end{equation}
Evaluating this at $s=\overline{s}=0$ implies that the closed superstring 
products are in the small Hilbert space and satisfy the $L_\infty$ relations. 

Now we have to solve \eq{Ldt}-\eq{Ldbs} to define the products. Expanding 
\eq{Ldt} in powers gives the formula
\begin{equation}\L_{n+2}^{(p,q)}=\frac{1}{p+q}\sum_{k=0}^n
\left(\sum_{r,s}[\L_{n-k+1}^{(r,s)},\ll_{k+2}^{(p-r,q-s)}]
+\sum_{r,s}[\L_{n-k+1}^{(r,s)},\llb_{k+2}^{(p-r,q-s)}]\right).\label{eq:IILrec}
\end{equation}
The sum over $r,s$ include all values such that the product and gauge product 
in the commutator have admissible picture numbers. Explicitly, in the 
commutator with $\ll$,
\begin{eqnarray}
\lineup \sup(0,p-k-1)\leq r\leq \inf(n-k,p-1),\nonumber\\
\lineup \sup(0,q-k-1)\leq s\leq \inf(n-k,q),
\end{eqnarray} 
and in the commutator with $\llb$,
\begin{eqnarray}
\lineup \sup(0,p-k-1)\leq r\leq \inf(n-k,p),\nonumber\\
\lineup \sup(0,q-k-1)\leq s\leq \inf(n-k,q-1).
\end{eqnarray}
Similar to \eq{Mrec}, this formula determines the products recursively given
the products of the bosonic string and the left/right gauge products.
The left/right gauge products are defined by solving \eq{Lds} and \eq{Ldbs},
and following the argument of section \ref{subsec:All} we find natural 
solutions 
\begin{eqnarray}
\l_{n+2}^{(p+1,q)}\lineup = \frac{n-p+1}{n+3}\Big(\xi_0 L_{n+2}^{(p,q)}
-L_{n+2}^{(p,q)}(\xi_0\w\mathbb{I}_{N+1})\Big),\label{eq:IIlrec}\\
\lb_{n+2}^{(p,q+1)}\lineup = \frac{n-q+1}{n+3}\Big(\overline{\xi}_0 
L_{n+2}^{(p,q)}-L_{n+2}^{(p,q)}(\overline{\xi}_0\w\mathbb{I}_{N+1})\Big).
\label{eq:IIlbrec}
\end{eqnarray}
Once we know all products and gauge products with up to $n+1$ inputs, 
we can determine the $(n+2)$nd superstring product $L_{n+2}^{(n+1,n+1)}$ 
by filling a ``diamond'' of products of intermediate picture number, starting
from the bosonic product $L_{n+2}^{(0,0)}$ at the bottom. Filling the 
diamond requires climbing $4(n+1)$ levels, $2(n+1)$ of those require 
computing gauge products from products using \eq{IIlrec} and \eq{IIlbrec}, 
and the other $2(n+1)$ require computing products from gauge products 
using \eq{IILrec}. 

Just to see this work, let's write the necessary formulas
to determine the 3-product $L_3^{(2,2)}$, corresponding to the ``diamond''
sketched in \ref{fig:diamond}. Start from the bosonic product:
\begin{equation}\L_3^{(0,0)}=\mathrm{given}.\end{equation}
In the first level we have two gauge products,
\begin{eqnarray}
\l_3^{(1,0)}\lineup =\frac{1}{2}\Big(\xi_0 L_3^{(0,0)}
-L_3^{(0,0)}(\xi_0\w\mathbb{I}_2)\Big),\\
\lb_3^{(0,1)}\lineup =\frac{1}{2}\Big(\overline{\xi}_0 L_3^{(0,0)}
-L_3^{(0,0)}(\overline{\xi}_0\w\mathbb{I}_2)\Big).
\end{eqnarray}
In the second level, two products:
\begin{eqnarray}
\L_3^{(1,0)} \lineup = [\Q,\ll_3^{(1,0)}]+[\L_2^{(0,0)},\ll_2^{(1,0)}],\\
\L_3^{(0,1)} \lineup = [\Q,\ll_3^{(0,1)}]+[\L_2^{(0,0)},\ll_2^{(0,1)}].
\end{eqnarray}
In the third level, four gauge products:
\begin{eqnarray}
\l_3^{(2,0)}\lineup =\frac{1}{4}\Big(\xi_0 L_3^{(1,0)}
-L_3^{(1,0)}(\xi_0\w\mathbb{I}_2)\Big),\\
\lb_3^{(1,1)}\lineup =\frac{1}{2}\Big(\overline{\xi}_0 L_3^{(1,0)}
-L_3^{(1,0)}(\overline{\xi}_0\w\mathbb{I}_2)\Big),\\
\l_3^{(1,1)}\lineup =\frac{1}{2}\Big(\xi_0 L_3^{(0,1)}
-L_3^{(0,1)}(\xi_0\w\mathbb{I}_2)\Big),\\
\lb_3^{(0,2)}\lineup =\frac{1}{4}\Big(\overline{\xi}_0 L_3^{(0,1)}
-L_3^{(0,1)}(\overline{\xi}_0\w\mathbb{I}_2)\Big).
\end{eqnarray}
In the fourth level, three products:
\begin{eqnarray}
\L_3^{(2,0)} \lineup = \frac{1}{2}
\Big([\Q,\ll_3^{(2,0)}]+[\L_2^{(1,0)},\ll_2^{(1,0)}]\Big),\\
\L_3^{(1,1)} \lineup = \frac{1}{2}
\Big([\Q,\ll_3^{(1,1)}+\llb_3^{(1,1)}]+[\L_2^{(0,0)},\ll_2^{(1,1)}
+\llb_2^{(1,1)}] 
+[\L_2^{(0,1)},\ll_2^{(1,0)}]+[\L_2^{(1,0)},\llb_2^{(0,1)}]\Big), \nonumber\\
\\
\L_3^{(0,2)} \lineup = \frac{1}{2}
\Big([\Q,\llb_3^{(0,2)}]+[\L_2^{(0,1)},\llb_2^{(0,1)}]\Big).
\end{eqnarray}
In the fifth level, four gauge products:
\begin{eqnarray}
\lb_3^{(2,1)}\lineup =\frac{1}{4}\Big(\overline{\xi}_0 L_3^{(2,0)}
-L_3^{(2,0)}(\overline{\xi}_0\w\mathbb{I}_2)\Big),\\
\l_3^{(2,1)}\lineup =\frac{1}{2}\Big(\xi_0 L_3^{(1,1)}
-L_3^{(1,1)}(\xi_0\w\mathbb{I}_2)\Big),\\
\lb_3^{(1,2)}\lineup =\frac{1}{2}\Big(\overline{\xi}_0 L_3^{(1,1)}
-L_3^{(1,1)}(\overline{\xi}_0\w\mathbb{I}_2)\Big),\\
\l_3^{(1,2)}\lineup =\frac{1}{4}\Big(\xi_0 L_3^{(0,2)}
-L_3^{(0,2)}(\xi_0\w\mathbb{I}_2)\Big).
\end{eqnarray}
In the sixth level, two products:
\begin{eqnarray}
\L_3^{(2,1)} \lineup = \frac{1}{3}
\Big([\Q,\ll_3^{(2,1)}+\llb_3^{(2,1)}]+[\L_2^{(1,0)},\ll_2^{(1,1)}
+\llb_2^{(1,1)}]+[\L_2^{(1,1)},\ll_2^{(1,0)}]\Big),
\\
\L_3^{(1,2)} \lineup = \frac{1}{3}
\Big([\Q,\ll_3^{(1,2)}+\llb_3^{(1,2)}]+[\L_2^{(0,1)},\ll_2^{(1,1)}
+\llb_2^{(1,1)}]+[\L_2^{(1,1)},\llb_2^{(0,1)}]\Big).
\end{eqnarray}
In the seventh level, two gauge products:
\begin{eqnarray}
\l_3^{(2,2)}\lineup =\frac{1}{4}\Big(\xi_0 L_3^{(1,2)}
-L_3^{(1,2)}(\xi_0\w\mathbb{I}_2)\Big),\\
\lb_3^{(2,2)}\lineup =\frac{1}{4}\Big(\overline{\xi}_0 L_3^{(2,1)}
-L_3^{(2,1)}(\overline{\xi}_0\w\mathbb{I}_2)\Big).
\end{eqnarray}
Finally, at the eighth level:
\begin{equation}\L_3^{(2,2)}=\frac{1}{4}\Big([\Q,\ll_3^{(2,2)}+\llb_3^{(2,2)}]
+[\L_2^{(1,1)},\ll_2^{(1,1)}+\llb_2^{(1,1)}]\Big),\end{equation}
which is the 3-product of the closed superstring.

Let us mention a few generalizations of this construction. 
Instead of \eq{Ldt}, we could define the products using the differential 
equation 
\begin{equation}\frac{\d}{\d t}\L(s,\overline{s},t)
=\Big[\L(s,\overline{s},t),c\,\ll(s,\overline{s},t)+\overline{c}\,
\llb(s,\overline{s},t)\Big],\label{eq:Lcombdt}
\end{equation}
for $c,\overline{c}$ arbitrary constants, while keeping equations \eq{Lds} 
and \eq{Ldbs} the same. It turns out that this setup can be transformed 
into the previous one by rescaling $\l,\lb$ and $s,\overline{s}$. The
resulting products are related by 
\begin{equation}L_{n+1}^{(p,q)}\,(\mathrm{derived\ from}\ \eq{Lcombdt})=
c^p\overline{c}^q L_{n+1}^{(p,q)}\,(\mathrm{derived\ from}\ \eq{Ldt})
\end{equation}
In particular, $L_{n+1}^{(n,n)}$ derived from \eq{Lcombdt} 
is related to $L_{n+1}^{(n,n)}$ derived from \eq{Ldt} by a trivial 
factor $(c\overline{c})^n$, which can be absorbed into a redefinition of
the coupling constant. A more nontrivial generalization is to 
take $c$ and $\overline{c}$ to be functions of $t$. This can be understood
as follows. The form of the generating functions \eq{NSNSgen} suggests that 
$\L,\ll$ and $\llb$ can be thought of as depending on a fourth variable 
$\overline{t}$, which counts the rank of ``rightmoving'' multiplication. 
However, since left and right multiplication is identified, $t$ and 
$\overline{t}$ are not independent variables, and in \eq{NSNSgen} we took 
$t=\overline{t}$. However, we can imagine a more general relation between 
$t$ and $\overline{t}$ where they are taken to be functions of an independent
parameter $\tau$. Then \eq{Ldt} is naturally generalizes to 
\begin{equation}\frac{\d}{\d \tau}\L(s,\overline{s},\tau)
=\left[\L(s,\overline{s},\tau),\frac{d t(\tau)}{d\tau}\,
\ll(s,\overline{s},\tau)+\frac{d\overline{t}(\tau)}{d\tau}\,
\llb(s,\overline{s},\tau)\right].\label{eq:curvedt}
\end{equation}
Note that the parameter $\tau$ does not (in general) count picture number, 
and the coefficients of a power series expansion of $\L(s,\overline{s},\tau)$ 
are general coderivations describing superpositions of the products with 
different picture numbers. This makes it difficult to extract the definition 
of the products from the solution to this differential equation. One 
application of this setup, however, is that the superstring products 
described here and those described in section \ref{subsec:comp} can be 
formulated in a common language. They follow from two different choices of 
curves in the $t,\overline{t}$ plane:
\begin{eqnarray}
\mathrm{This\ section:}\lineup\ \ \ t(\tau)=\tau,\ \ \overline{t}(\tau)=\tau,\\
\mathrm{Section}\ \ref{subsec:comp}:\lineup\ \ \ \left\{\begin{matrix}
t(\tau)=\tau,\ \ \ \overline{t}(\tau)=0,\ \ \ \ \ \ \ \ 
\tau\in[0,T],\ \ \\
\,t(\tau)=T,\ \ \ \overline{t}(\tau) = \tau-T,\ \ \tau\in[T,2T].
\end{matrix}\right.
\end{eqnarray}
In the former case, the products follow by evaluating $\L$ at 
$s=\overline{s}=0$ and $\tau=T$ and 
expanding in powers of $T$, while in the latter case, 
they follow from evaluating $\L$ at $s=\overline{s}=0$ and 
$\tau=2T$ and expanding in powers of $T$. This gives one possible avenue to
the proof of gauge equivalence between the products derived here and in 
section \ref{subsec:comp}.

\section{Conclusion}

In this paper we have constructed explicit actions for all NS superstring 
field theories in the small Hilbert space. Closely following the calculations 
of \cite{WittenSS}, one can show that they reproduce the 
correct 4-point amplitudes. Since these actions share the same
algebraic structure as bosonic string field theory, relaxing the ghost 
number of the string field automatically gives a solution to the classical 
BV master equation. This is a small, but significant step towards the goal
of providing an explicit computational and conceptual understanding 
of quantum superstring field theory. The next steps of this program include
\begin{itemize}
\item Incorporate the Ramond sector(s) so as to maintain a controlled 
solution to the classical BV master equation.
\item Quantize the theory. Specifically determine the higher genus 
corrections to the tree-level action needed to ensure a solution to the 
quantum BV master equation.
\item Understand how the vertices and propagators of classical or 
quantum superstring field theory provide a single cover of the supermoduli 
space of super-Riemann surfaces.
\item Understand how this relates to formulations of superstring field theory
in the large Hilbert space, which may ultimately be more fundamental.
\end{itemize}
Progress on these questions will not only help to assess whether superstring
field theory can be a useful tool beyond tree level, but may 
provide valuable insights into the systematics of superstring 
perturbation theory.

\bigskip

\noindent {\bf Acknowledgments}

\bigskip

\noindent This project was supported in parts by the DFG Transregional 
Collaborative Research Centre TRR 33, the DFG cluster of excellence Origin and Structure of the Universe. 

\begin{appendix}

\section{Quartic Vertices}
\label{app:vert}

The recursive construction of the vertices ultimately defines the action 
in terms of $X$s, $\xi$s, and products of the bosonic string. However, it 
is not necessarily easy to derive an explicit expression for the action 
in this form. The coalgebra notation offers great notational efficiency in 
expressing the recursive definition of the products, but it does not directly 
display the cyclically inequivalent contributions to each vertex (or, for 
the case of the closed string, the symmetrically inequivalent contributions). 
To obtain the cyclically or symmetrically inequivalent contributions, one 
must solve the recursion to the relevant order, expand the the multi-string 
product (term-by-term) into $X$s, $\xi$s, and bosonic products, and then place 
each term into the respective cyclic or symmetric equivalence class. This
 procedure quickly becomes impractical to implement by hand; for example, 
the NS-NS quartic vertex involves 91 symmetrically inequivalent 
contributions, though depending on the construction some terms may vanish 
or be related by left/right symmetry. However, we are able to execute the 
computation out to quartic order for the open and heterotic string. We
present our results here. It is an important open question whether a more 
efficient or direct method for computing the vertices in this form can be 
obtained.

Since individual contributions to each vertex contain the $\xi$ zero mode, 
it is convenient to write the action using the symplectic form in 
the large Hilbert space. This is related to the symplectic form in the 
small Hilbert space through the formula \cite{WittenSS}
\begin{equation}\langle \omega_L|(\mathbb{I}\otimes \xi)(L\otimes L)
=\langle\omega|, \end{equation}
where $L$ is the trivial map from the small Hilbert space to $\eta$-closed 
elements in the large Hilbert space. The vertices in the action 
can be written in several equivalent forms, related by cyclicity or by 
$\eta$-exact contributions. We fix this redundancy by requiring that one 
factor of $X$ in the vertex always appears multiplied by $\xi$, and if 
$\xi X$ acts on an external state, it always acts on the first input of the 
symplectic form. If $\xi X$ does not act on an external state, there is 
always a remaining $\xi$ which can again be chosen to act on the first 
entry of the symplectic form. With these choices, the cubic vertex in the 
open superstring action is 
\begin{equation}\frac{1}{3}\omega\left(\Phi,M_2^{(1)}(\Phi,\Phi)\right)=
\frac{1}{3}\omega_L\left(\xi X\Phi,M_2^{(0)}(\Phi,\Phi)\right).\end{equation}
The quartic vertex takes the form:
\begin{eqnarray}
\frac{1}{4}\lineup \omega\left(\Phi,M_3^{(2)}(\Phi,\Phi,\Phi)\right) = 
\nonumber\\
\lineup\ \ \ \ \ \frac{5}{36}\ \left[
\omega_L\left(\xi X \Phi,M_2^{(0)}(\Phi,\xi M_2^{(0)}(\Phi,\Phi))\right)
+\omega_L\left(\xi X \Phi,M_2^{(0)}(\xi M_2^{(0)}(\Phi,\Phi),\Phi)\right)
\right]\nonumber\\
\lineup\ \ +\frac{1}{144}\left[
\omega_L\left(\xi X \Phi,M_2^{(0)}(\Phi, M_2^{(0)}(\xi\Phi,\Phi))\right)
+\omega_L\left(\xi X \Phi,M_2^{(0)}(\Phi, M_2^{(0)}(\Phi,\xi\Phi))\right)
\right.
\nonumber\\
\lineup\ \ \ \ \ \ \ \ \ \,\left.
+\omega_L\left(\xi X \Phi,M_2^{(0)}(M_2^{(0)}
(\xi\Phi,\Phi),\Phi)\right)
+\omega_L\left(\xi X \Phi,M_2^{(0)}(M_2^{(0)}(\Phi,\xi\Phi),\Phi)\right)
\right]\nonumber\\
\lineup\ \ +\ \frac{1}{36}\ \left[
\omega_L\left(\xi \Phi,M_2^{(0)}(\Phi,\xi X M_2^{(0)}(\Phi,\Phi))\right)
+\omega_L\left(\xi \Phi,M_2^{(0)}(\xi X M_2^{(0)}(\Phi,\Phi),\Phi)\right)
\right]\nonumber\\
\lineup \ \ +\ \frac{1}{16}\ \left[
\omega_L\left(\xi X \Phi,M_2^{(0)}(M_2^{(0)}(\Phi,\Phi),\xi\Phi)\right)
-\omega_L\left(\xi X \Phi,M_2^{(0)}(\xi\Phi, M_2^{(0)}(\Phi,\Phi))\right)
\right]\nonumber\\
\lineup\ \ +\ \frac{1}{16}\ \left[
\omega_L\left(\xi X^2 \Phi,M_3^{(0)}(\Phi,\Phi,\Phi)\right)
+\omega_L\left(\xi X \Phi,M_3^{(0)}(\Phi,X\Phi,\Phi)\right)
\right]\nonumber\\
\lineup \ \ +\ \ \frac{1}{8}\ \,\left[
\omega_L\left(\xi X \Phi,M_3^{(0)}(X\Phi,\Phi,\Phi)\right)
\right].
\end{eqnarray}
If $M_2^{(0)}$ is Witten's associative star product, we can set 
$M_3^{(0)}=0$ and the quartic vertex simplifies to:
\begin{eqnarray}
\frac{1}{4}\lineup \omega\left(\Phi,M_3^{(2)}(\Phi,\Phi,\Phi)\right) = 
\nonumber\\
\lineup\ \ \ \ \frac{5}{36}\left[
\omega_L\left(\xi X \Phi,M_2^{(0)}(\Phi,\xi M_2^{(0)}(\Phi,\Phi))\right)
+\omega_L\left(\xi X \Phi,M_2^{(0)}(\xi M_2^{(0)}(\Phi,\Phi),\Phi)\right)
\right]\nonumber\\
\lineup\ \ -\frac{1}{18}\left[
\omega_L\left(\xi X \Phi,M_2^{(0)}(\Phi, M_2^{(0)}(\Phi,\xi\Phi))\right)
+\omega_L\left(\xi X \Phi,M_2^{(0)}(\xi\Phi, M_2^{(0)}
(\Phi,\Phi)\right)\right]\nonumber\\
\lineup\ \ +\frac{1}{36}\left[
\omega_L\left(\xi \Phi,M_2^{(0)}(\Phi,\xi X M_2^{(0)}(\Phi,\Phi))\right)
+\omega_L\left(\xi \Phi,M_2^{(0)}(\xi X M_2^{(0)}(\Phi,\Phi),\Phi)\right)
\right].\nonumber\\
\end{eqnarray}
The 3-vertex for the heterotic string is
\begin{equation}\frac{1}{3!}\omega\left(\Phi,L_2^{(1)}(\Phi,\Phi)\right)=
\frac{1}{3!}\omega_L\left(\xi_0 X_0\Phi,L_2^{(0)}(\Phi,\Phi)\right),
\end{equation}
and the 4-vertex is
\begin{eqnarray}
\frac{1}{4!}\lineup \omega\left(\Phi,L_3^{(2)}(\Phi,\Phi,\Phi)\right) = 
\nonumber\\
\lineup\ \ \ \ \ \ \frac{5}{108}
\omega_L\left(\xi_0 X_0 \Phi,L_2^{(0)}(\Phi,\xi_0 L_2^{(0)}(\Phi,\Phi))\right)
+\frac{1}{216}\omega_L\left(\xi_0 X_0 \Phi,L_2^{(0)}(\Phi, L_2^{(0)}(\Phi,
\xi_0\Phi))\right)
\nonumber\\
\lineup\ \ \ +\ \frac{1}{108}\omega_L\left(\xi_0 \Phi,L_2^{(0)}(\Phi,\xi_0 X_0 
L_2^{(0)}(\Phi,\Phi))\right)-\frac{1}{48}
\omega_L\left(\xi_0 X_0 \Phi,L_2^{(0)}(\xi_0\Phi, L_2^{(0)}(\Phi,\Phi))\right)
\nonumber\\
\lineup\ \ \ +\ \frac{1}{96}
\omega_L\left(\xi_0 X_0^2 \Phi,L_3^{(0)}(\Phi,\Phi,\Phi)\right)
+\frac{1}{32}\omega_L\left(\xi_0 X_0 \Phi,L_3^{(0)}(\Phi,\Phi,X_0\Phi)\right).
\end{eqnarray}

\end{appendix}

\end{document}